# Quantum kernels for real-world predictions based on electronic health records

Zoran Krunic[1†], Frederik F. Flöther[2], George Seegan[1], Nathan Earnest-Noble[3], Omar Shehab[3]

[1] Amgen, One Amgen Center Drive, Thousand Oaks, CA, 91320-1799, USA
[2] IBM Quantum, IBM Switzerland Ltd, Vulkanstrasse 106, CH-8010 Zurich, Switzerland
[3] IBM Quantum, IBM Thomas J. Watson Research Center, 1101 Kitchawan Rd, Yorktown Heights, NY 10598, USA

**Abstract**

In recent years, research on near-term quantum machine learning has explored how classical machine learning algorithms endowed with access to quantum kernels (similarity measures) can outperform their purely classical counterparts. Although theoretical work has shown provable advantage on synthetic data sets, no work done to date has studied empirically whether quantum advantage is attainable and with what kind of data set. In this paper, we report the first systematic investigation of empirical quantum advantage (EQA) in healthcare and life sciences and propose an end-to-end framework to study EQA. We selected electronic health records (EHRs) data subsets and created a configuration space of 5–20 features and 200–300 training samples. For each configuration coordinate, we trained classical support vector machine (SVM) models based on radial basis function (RBF) kernels and quantum models with custom kernels using an IBM quantum computer, making this one of the largest quantum machine learning experiments to date. We empirically identified regimes where quantum kernels could provide advantage on a particular data set and introduced a terrain ruggedness index, a metric to help quantitatively estimate how the accuracy of a given model will perform as a function of the number of features and sample size. The generalizable framework introduced here represents a key step towards a priori identification of data sets where quantum advantage could exist.

**Introduction**

Over the last years, real-world data has been increasingly used to generate medical evidence and progress precision medicine. This includes sources such as electronic health records (EHRs), claims and billing data, product and disease registries, and data from wearables and health applications[1]. Powerful data mining techniques have been applied to such data sets, particularly to EHRs, in order to predict a broad range of medical conditions and events[2,3,4,5]. However, classical machine learning and data science techniques have limitations with regard to learning some of the most complex patterns; for instance, the predictive power of genetic risk scores derived from genome-wide association studies has plateaued over the last years[6]. As a result, quantum machine learning has been explored as an alternative and for certain general problems it has already been proved that quantum machine learning algorithms can provide benefits beyond the scope of classical ones[7]. The complexity of the correlations and patterns in EHRs and (real-world) medical data sets makes such data sources prime candidates for the application of quantum algorithms[8].

† Electronic address: zkrunic@amgen.com

The study of supervised machine learning problems with quantum techniques is an active area of research[9]. In early work on classification with near-term quantum algorithms[10,11,12], the proposed quantum feature maps typically encode the datapoints into inner products or amplitudes in the Hilbert space. The quantum circuit used to implement the feature map is of a length which is typically a linear or polylogarithmic function of the size of the data set and the number of qubits is a function of the number of features. In subsequent work, the advantage of a quantum feature map was rigorously proved for a carefully chosen synthetic data set[7]. Recently, a body of work[13,14] implementing quantum feature maps for small-scale coarse-grained practical data sets has emerged; while there have been studies of different feature maps[15], none have been discovered so far with rigorous advantage for general or practical data sets. The capability limits of near-term quantum computers has also been pushed in work where the data set was less coarse-grained[16,17]; furthermore, efforts have begun to study how hyperparameters affect the potential advantage of a given quantum classifier[18]. Another stream of research has emerged on finding a suitable quantum feature map for a given data set with [19] providing a recent review of quantum classification algorithms. Studies on quantum feature maps involve both the study of the kernel function and the study of the quantum circuits which encode the outcome of the kernel function into Hilbert space. A new set of metrics and a protocol has also been proposed to determine the possibility of quantum advantage for a given pair of data set and quantum feature map[20].

In this work, we focus on one kernel-based method which uses the quantum support vector machine (QSVM), estimating the kernel with a quantum computer and feeding it back into a classical support vector machine (SVM) for classification[12]. To the best of our knowledge, there have not yet been any systematic studies regarding the applicability of quantum kernels to EHRs. Here we predict the six-month persistence of rheumatoid arthritis patients on biologic therapies. The central research questions investigated in this work are therefore:
- Can we enhance the prediction of medication persistence by applying quantum kernels to real-world EHRs?
- Can we systematically identify problem instances (number of features, number of samples) where quantum computing may have an advantage for such real-world data sets?

The methods developed in this work are general and can be applied for a wide variety of problems with different-size data sets in machine learning and optimization. In this paper, nevertheless, we focus on small data sets, particularly those where the ratio of the number of features to the number of samples is relatively large, which typically engender hard classification problems. Such data sets are important in a range of medical settings, for instance in clinical trials, studies of very specific cohorts, and translational medicine. Moreover, small data sets are naturally suited to near-term quantum computers.

The concepts "quantum supremacy"[21] and "quantum advantage"[22] have been around for a while and refer to asymptotic performance comparisons between a quantum approach and the best classical approach. Complementing these two foundational concepts, in this work we introduce a related concept called empirical quantum advantage (EQA). We define EQA as the incremental gain of using a specific quantum approach over a specific classical approach for a given problem. Once this heuristic measure is calculated, it is meaningful only in the context of three elements – the problem as well as the classical and quantum approaches used. It may not give any general asymptotic indications about "supremacy" or "advantage" for a family of problems. However, as in the field of practical classical algorithms[23], practitioners may use EQA to observe trends in empirical data. This is key in biology and medicine where both theoretical and operational factors must be considered, in general, when exploring the benefits of quantum algorithms for a given application[24].

When making measurements of EQA, multiple metrics were considered, with a final choice of three key metrics – F1 score and balanced accuracy at the configuration space coordinate level as well as the phase space terrain ruggedness index (PTRI) at the configuration space landscape level. PTRI is thus a global metric, fully described in the Methods section. The reasoning behind choosing these metrics was as follows.

Both F1 score and balanced accuracy are commonly used in machine learning; they measure the performance of a given model. On the other hand, PTRI captures the hardness of the configuration space for a given set of machine learning problems. The typical coordinate structure to explore that space of problems consists of the number of features and the number of samples. While the final data set used for binary classification is quite balanced (52 % to 48 %), more imbalanced cases were also considered. F1 score and balanced accuracy are thus readily generalizable to more imbalanced settings in future research. Furthermore, we chose to present the F1 score because, while it equally weights false negatives and false positives, we do not have the exact cost of either of those. In other words, the relative cost of recall and precision are different in specific model deployments.

**Results**

We started by evaluating multiple two-dimensional landscapes in the classical domain with the number of topmost important features ranging from 1 to 20 in increments of 1 and the training set sizes ranging from 50 to 600 samples in increments of 50. The topmost features were determined using the SHAP method[25] (see Methods section). For each landscape coordinate, 200 random train/test subsets were created out of the available data. Since the most resource- and time-demanding part of the process is calculation of the custom kernels in quantum simulations (classical hardware simulating the behavior of a quantum computer) and on a real quantum processing unit (QPU), a small number of these data sets were chosen to evaluate the required total processing time. Custom kernels were calculated using both quantum simulations and QPUs. With the obtained runtimes, a realistic number of data sets that could be executed was calculated. As a result, the configuration landscape was reduced to four feature number values [5, 10, 15, 20] and three training set sizes [200, 250, 300], yielding a 12-point configuration space. For each configuration space coordinate, two data sets, each containing a train and a test subset, were selected out of the 200 train/test subsets. This was achieved by calculating balanced accuracy in the classical domain and then selecting two sets with the balanced accuracy close to the mean of the balanced accuracies of the full 200-sample set. The term subpoint is used to denote each individual data set within the given coordinate.

This yielded a total of 24 subpoints across the 12-coordinate configuration space. We found this to be feasible for quantum simulation and, critically, also for QPU execution. Since analytical study of the hardness of such a large practical problem is extremely difficult, these types of large-scale simulation and hardware experiments across a broad configuration space are the most pragmatic way to identify trends and outliers. As a parallel outcome, this work hence represents one of the largest quantum machine learning experiments to date. The feature size component of the coordinates dictates the number of qubits for the QSVM. It was obtained by taking the most important features from the classical models built on the same data using the full-size data sets (see Methods section).

We used the predict method from the svm.SVC class within scikit-learn[26] as the main method for comparing quantum and classical support vector machine performance. In addition, the predict_proba method was used to obtain estimates of probabilities. Thresholds were varied in the range 0 to 1 in small increments and applied to the probabilities to generate the optimal split between the two class labels. The plots presented focus on the predict method; a detailed discussion of the probability-based results can be found in the Methods section.

Presented in Fig. 1 are the comparative 3D plots of F1 score and balanced accuracy metrics with the orange surface presenting points of the averaged metric for classical computing and the blue surface for quantum computing (QPU). All QPU experiments presented in this paper were run on ibmq_dublin (see Methods section). Each point on the configuration space coordinate was averaged from two selected data sets for that coordinate; thus, each plotted configuration space has 12 points in total. The z-axis is the metric while the x- and y-axis are the number of features and training samples respectively.

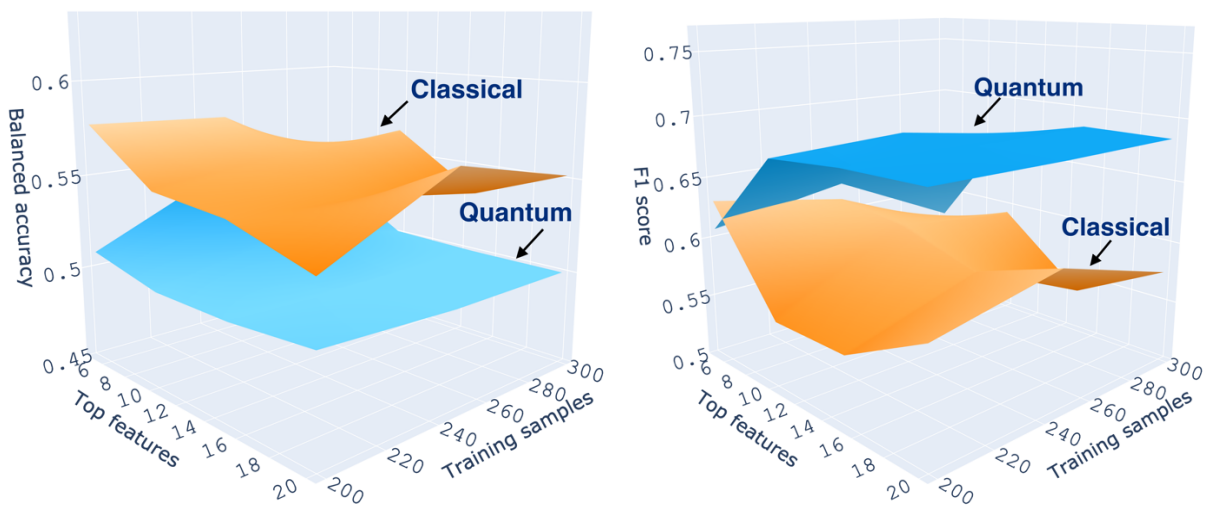

Fig. 1. Left: Balanced accuracy landscape. Right: F1 score landscape. The classical domain is shown in orange and the quantum domain in blue.

The PTRI was calculated for the full configuration space both for balanced accuracy and F1 score metrics and plotted in Fig. 2 using the same approach as previously presented for the other metrics.

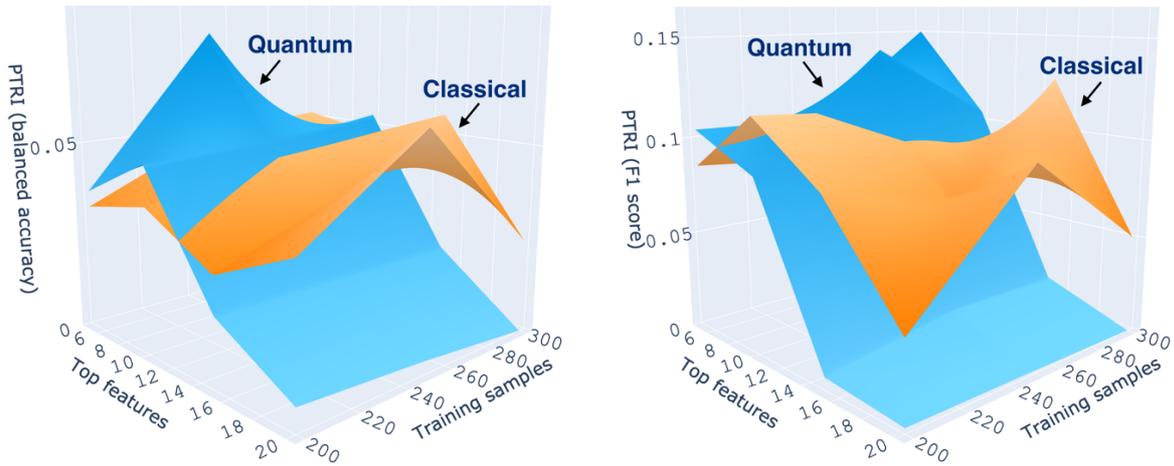

Fig 2. Left: PTRI (balanced accuracy) landscape. Right: PTRI (F1 score) landscape. The classical domain is shown in orange and the quantum domain in blue.

For each point in the configuration space and its corresponding two subpoints, we calculated the geometric difference defined in [20] between radial basis function (RBF) and quantum kernels and averaged the corresponding two values at each coordinate, as shown in Fig. 3.

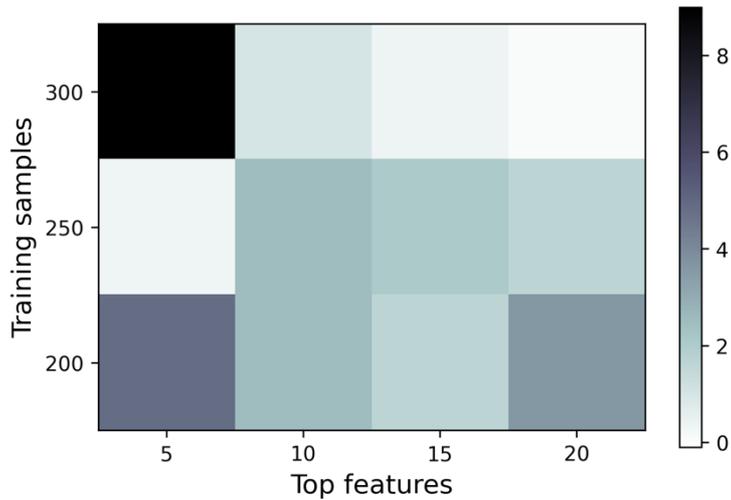

Fig. 3. Geometric difference between classical and quantum kernels across the configuration space. Despite its name, the mathematical definition demonstrates that it is a ratio. The greater the geometric difference, the more potential for quantum advantage the given quantum kernel has compared to the given classical kernel.

The plots in Fig. 4 illustrate the results obtained with QPU and classical processing for different metrics. The plots show the difference of the metric value achieved by QPU minus the metric value calculated by classical SVM. Since there are two random data sets for each coordinate, those are averaged and then the difference is calculated, resulting in 12 configuration space points. The points are plotted as three lines, one for each training set size, across four feature number values.

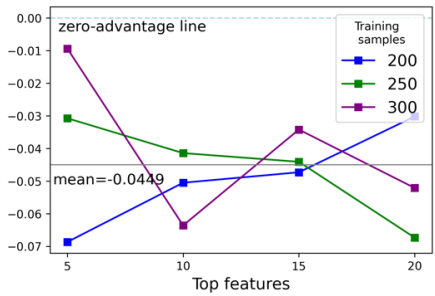 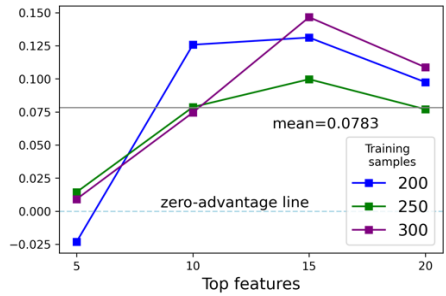
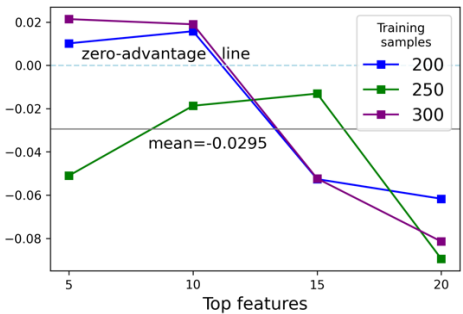 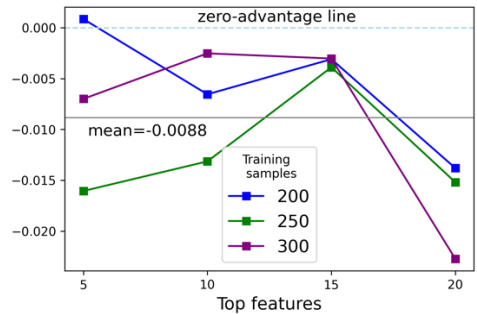

Fig. 4. QPU balanced accuracy and F1 score vs. classical balanced accuracy and F1 score, non-probability and probability-based approach.

By comparing the position of the data points with the horizontal "zero-advantage line", we observe EQA for a subset of problem instances in the configuration space: 0 % and 92 % of all instances for non-probability-based balanced accuracy and F1 score respectively as well as 33 % and 8 % of all instances for probability-based balanced accuracy and F1 score respectively showed such quantum advantage.

**Discussion**

In this work, we have considered a configuration space of classification problems with varying numbers of features and samples. On that manifold, we have observed EQA for 0 % and 92 % of classification problem instances for non-probability-based balanced accuracy and F1 score respectively as well as 33 % and 8 % of classification problem instances for probability-based balanced accuracy and F1 score respectively. This makes it apparent that EQA is something that must be evaluated on a case-by-case basis until clearer trends present themselves. Identification of hard instances through careful domain consideration has allowed us to observe such advantages with no circuit being executed more than 1024 times (i.e. a maximum of 1024 shots), which is almost one order of magnitude less than previous large-scale quantum machine learning experiments[16,17] (and therefore results in higher sampling noise). This indicates that, in the future, domain expertise about the hardness of practical problems is going to be crucial for the development and refinement of quantum algorithms. Our observation of these empirical trends reiterates the significance of developing such large-scale experiments to understand the trends and detect outliers. The considerable differences in EQA based on the choice of performance metric could suggest that practical quantum advantage is going to be highly domain-specific. Further work is needed to explore the applicability of different performance metrics for various domains. Given that there is a need to build robust machine learning models in medical settings where additional samples are costly or impossible to acquire, even a modest reduction in the number of samples required for training based on certain data distributions can yield considerable benefits for many prediction and inference problems in biology[24].

We also introduced a practical metric, PTRI, to quantify and thereby qualify the quantum advantage potential for a given problem. For any metric, PTRI helps identify the flattest and most rugged regions in configuration space. One could imagine that the flattest classical performance region is the configuration subspace where the performance of the classical techniques becomes stagnant and where a quantum algorithm should therefore be considered. In that case, computing the PTRI for the quantum approach over the given configuration space may give some insights about where quantum advantage is likeliest. This domain-agnostic metric is one of the first attempts for an operational tool which, in the future, quantum practitioners can use to determine when to use a quantum computer, a dynamic decision that may have to be taken very frequently, under severe timing constraints. Further study is needed to interpret the amount of correlation between the manifolds of classical and quantum performance metrics in terms of PTRI and related measures.

As a parallel result, we have also presented, to the best of our knowledge, the first independent application of the geometric difference, which we employed to determine the relative separation between classical and quantum feature maps. Further study is needed to understand how quantum practitioners may combine the concepts of PTRI and geometric difference to first identify the potential for quantum advantage in a configuration subspace and then estimate the potential of a specific quantum feature map in that subspace. We emphasize that there may be other relevant metrics worth exploring in the future when studying forms of quantum advantage, such as energy consumption.

It is also important to observe that we used the same kernel function and feature map for every classification problem. More studies are needed to determine appropriate combinations of kernel function and feature map that result in greater EQA. It may also be worthwhile to investigate whether there are kernel functions inspired by one-way[7], trapdoor, or learning with error (LWE)[27] protocols that may not only provide advantage in prediction accuracy but also in time complexity.

Ultimately, we conducted the first systematic study of QSVM configuration space and quantum classification based on an EHR data set. We classified the persistence of rheumatoid arthritis patients on biologic therapies, predicting six-month persistence via binary classification. Furthermore, we proposed an end-to-end framework to study EQA that can be generalized for other machine learning and optimization problems and observed EQA for a subset of problem instances in the configuration space. Our framework represents progress towards a priori identification of data sets where quantum advantage could be achieved and underscored that even with current quantum computers it is possible to arrive at predictions which are at least as good as those obtained with classical computers. These results have implications for classification problems across industries, particularly for small data sets.

# Methods

*Quantum Feature Map*

The feature map used in this work is known as the ZZFeatureMap, which gives rise to a feature space of $2^N$ dimensions where N is the number of qubits[12]. This family of circuits is believed to be hard to simulate classically[28].

*IBM Quantum hardware*

ibmq_dublin is a 27-qubit superconducting qubit quantum computer available on the IBM Quantum Services. The qubit connectivity is shown in Fig. 5. For qubits, lighter color means higher T2 time, and for couplings, lighter color means lower fidelity.

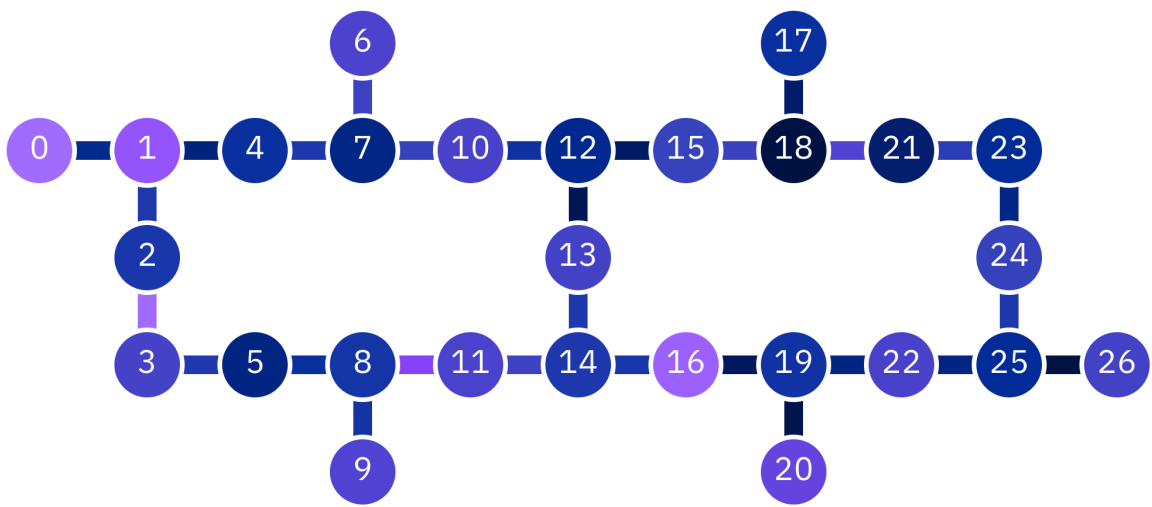

Fig. 5. Qubit connectivity of ibmq_dublin. For qubits, lighter color means higher T2 time; for couplings, lighter color means lower fidelity.

The average CNOT error rate and average readout error rate, at the time of authoring this manuscript, were 1.097 %, and 3.585 % respectively. The average T1 and T2 times were 107.03 μs and 114.53 μs respectively. The average gate time was 473.397 ns. More details may be accessed in real time[29]. For every quantum circuit, 1024 shots were run. The circuits were always maximally optimized using application programming interface (API) calls before the runs. A sample circuit computing the feature map of a five-feature data set is given in Fig. 6.

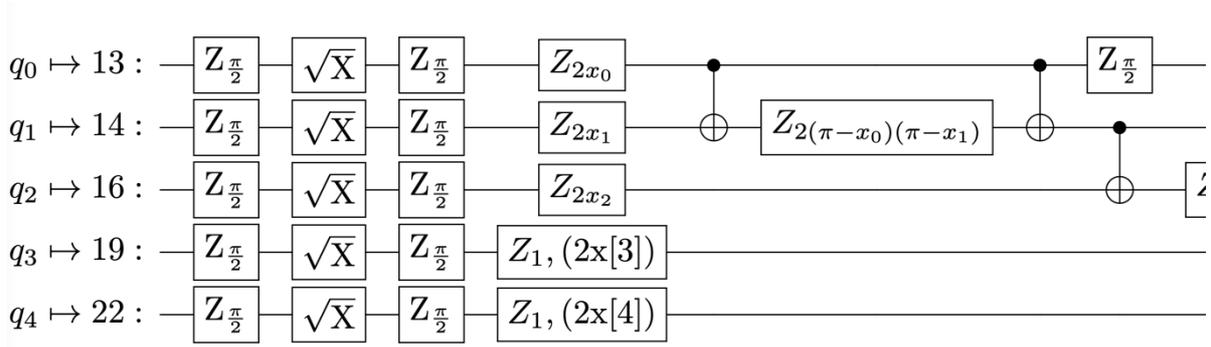

Fig. 6. Cropped circuit for quantum kernel calculations with full subcircuit feature for one inner product calculation on the top two qubits for a five-feature instance. The full circuit repeats a similar pattern across different qubit pairs. Qubits and interactions were mapped based on the connectivity of ibmq_dublin. $q_i$ is the i-th algorithmic qubit and the double-digit index after the arrow sign is the physical qubit index on ibmq_dublin.

*Quantum simulator*

The quantum simulations were run without noise models on the qasm_simulator, available on the IBM Quantum Services. Each circuit was run with 1024 shots and the circuits were always maximally optimized before each run. The simulations supported the experimental design and results.

*EHR data*

In this work, there have been two main challenges related to making predictions based on the EHR data. First, the problem of binary classification in patient persistence depends on the quality of the main classification label – that is, whether or not a given patient is persistent on the medication. This is derived from prescription, and there are known challenges in determining patient persistence from prescriptions[30,31]. While additional claims-based data sets could be used in conjunction with EHR data to improve the certainty of prescription patterns, that was not an available option during this work.

In addition, we used imputation to fill in missing data points. For example, not all laboratory results are equally present in each of the patients, and more specifically, not in the period of time covered by the data used for model training and testing. As detailed in the data section, the data was chosen such that for the first 10 features there is no missing data for any selected patient. Above the 10[th] feature, the data is sparser. Thus, for the features 11-20 missing data was imputed using the mean of the present data. While the impact of that imputation is minimal given that the majority of the model accuracy comes from the top 10 features, applying different imputation techniques could be explored in the future.

*Cohort restrictions*

The models were built using data from the Optum® Electronic Health Record data set, which includes deidentified and aggregated clinical and medical administrative data from over 100 million longitudinal EHR lives. Fields that were used included demographics, laboratory tests, observations, prescriptions, visits, and selected subsets of extracts from physicians' notes pre-processed using natural language processing (NLP) methods. A list of RA

(Rheumatoid Arthritis) International Classification of Diseases (ICD) diagnosis codes were used to select a first set of patients, further narrowed down by the given biologic's National Drug Codes (NDCs). The persistence for a given patient was defined as the length of time from initiation to discontinuation of the biologic therapy.

The therapy start date, i.e. the index date, was set to be the start date of the first biologic prescription. Inclusion criteria required at least one year of data prior to the index date; any data prior to one year before the index date was truncated, thus guaranteeing the same interval length for all patients. Additional inclusion criteria applied were a minimum of 6 months of data after the index date with stable payer insurance and the patient had to be at least 18 years of age and be in an integrated delivery network (which was indicated by the flag in the data set). Exclusion criteria were more than one diagnosis of systemic lupus erythematosus (SLE) or psoriatic arthritis (PsA), combined with prior use of targeted disease-modifying antirheumatic drugs (DMARDs) including biologics and Janus kinase inhibitors (JAKs). The inclusion/exclusion process is illustrated in Fig. 7.

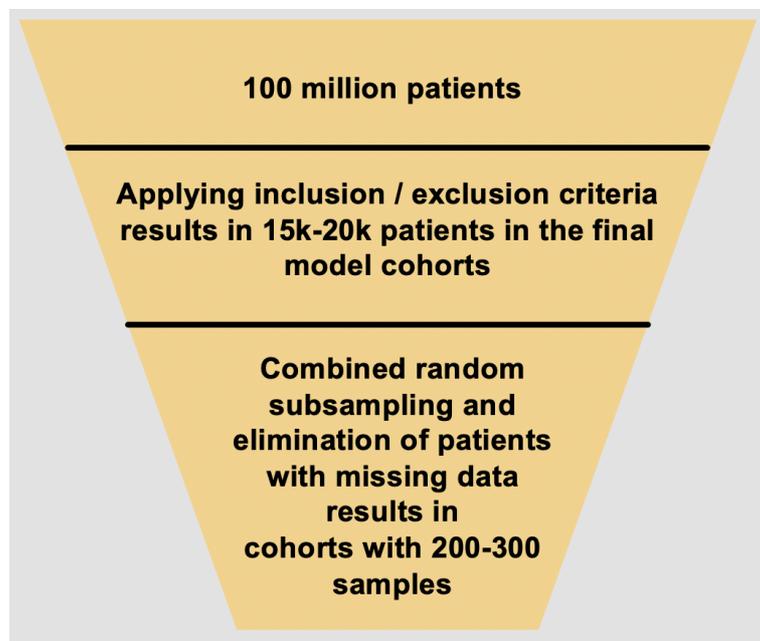

Fig. 7. Cohort restrictions applied to the EHR data.

*Creation of train and test subsets*

The full pipeline was developed on AWS/Databricks using PySpark, Python, and SQL. At the point where all preprocessing, inclusion and exclusion criteria were completed and the model training started, the data set size was reduced to 16000 samples, with a relatively balanced target class (52 % of patients persisted while 48 % did not). The number of features in that model exceeded 500, with the top 10 features accounting for more than 90 % of the achieved accuracy of 0.64. The variability of the model metrics within the set of 10 different train/test splits was under 4 %. The train/test split ratio was set at 80/20.

In the preliminary experiments with classical SVMs and quantum simulations, the range of the explored landscape was 50-600 in training set size in increments of 50 and the range in the number of features was from 1 to 20 in increments of 1. Since it was known that the top

20 features carry > 95 % of accuracy, this was deemed sufficient and within the reach of QSVM, where each feature maps to one qubit, thus leading to 20-qubit QPU experiments.

To reduce the training set size further from 12800 (80 % of the original 16000), an additional step was applied to the 16000-sample model data. First, the samples with no missing data in the top 10 features were selected, leaving the features from 11 to 20 with some missing data. That reduction yielded a data set of 1300 samples. This resulted in data sets with minimum missing data while preserving the top 20 features required for the experiments. Since not all patients have all the selected laboratory measurements or other features collected during the year before the start of the medication, we used imputation to fill in those values. Using a longer period of 2-3 years of training data prior to the index date significantly increases the chance of a patient having at least some value for the given features but reduces the overall number of patients in the cohort; therefore, that approach was not utilized in the final model.

The size of the training set was narrowed down to three different values – 200, 250, and 300. The final choice was to use 5, 10, 15, and 20 features, that, when combined with 200, 250, and 300 training set sizes, yielded a 12-point configuration space. From the 1300-sample data set, random sampling was used to create training data sets of 200, 250, and 300 samples. The test set size was kept at 150 to balance the constraints of reasonable runtimes for simulations and QPU while achieving the best stratification of samples under the circumstances.

The downside of the training (and test) data set reduction to a few hundred samples is the reduction in predictive accuracy, originally at 0.64 with 16000 samples. This decision was made in order to explore the very difficult cases where it is hard to get predictions better than random guesses. While we could have chosen starting models with 15000–20000 samples and accuracies above 0.75, which were available with somewhat different patient cohort structures, our goal was to tackle the most difficult problems. This careful consideration in selecting harder instances has allowed us to observe empirical advantage of quantum kernels over classical kernels even though none of the circuits was run with more than 1024 shots. In future work, and as quantum hardware and software scales further, we would like to explore train/test data sets with 500-1000 samples, which would allow for reduced loss of accuracy and smaller variability due to reduction in set size.

Knowing that there will be significant variability in the performance of different train/test data sets for each coordinate due to the small sample size, the maximum possible number of sets was evaluated. Given the preliminary runtimes for quantum simulations and QPU, the decision was made to use two random train/test splits for each configuration space coordinate, resulting in a total of 24 data sets to be run on QPU. While two random splits for each coordinate do not fully account for the variability resulting from such small data sets, this had to be limited due to QPU availability and simulation runtimes. Future work could be done to increase the number of data sets for each coordinate from two to 10 or more.

During subsampling it was ensured that the target classification label proportion was kept in the original proportion within each train and test subset. Different models and class imbalance ratios ranging from 1:1 to 1:5 were evaluated and the final decision was to use the aforementioned (almost) balanced class to reduce the impact of small data set size. It was our judgement that more imbalanced cases would be better addressed in subsequent research.

One of the main challenges with the small data sets is that when splitting train and test sets and training models on multiple splits, the resulting model metrics vary widely, especially with the models where predictive accuracy is not very high. To explore that, 200 random splits at each of the 12 coordinates were made. We calculated classical SVM balanced accuracies for each of the splits, a process that executed in less than one hour.

For each of the 24 subpoints, having already calculated classical SVM metrics, quantum simulations and QPU runs were executed, both using the Qiskit framework. 1024 shots were used as that allowed the execution for all 24 points within the time and resources available. The simulations were run with callback specified to provide additional insight during the running processes. The optimization level was set at three, the feature map to the ZZFeatureMap, the feature dimension equal to the number of features for the specific subpoint, the number of reps equal to two, and the entanglement to linear.

Quantum simulation and QPU processing were used to calculate custom kernel matrices for the given train and test set. The train kernel was then saved and used in scikit-learn on the classical computer to train the SVM model using a precomputed option. The test kernel was passed to the model's predict method to make predictions. This way, we generated predictions for quantum simulations and QPU runs. The classical predictions were generated using kernels in SVMs. The models were trained for 18 different values of the regularization parameter C, ranging from 0.006 to 1024, and the best case was used from each model for the classical to quantum comparison. Every model was regularized separately for each of the two metrics; thus, the value of the optimal C parameter for a given set's balanced accuracy is generally different than the value of C for the F1 score. All three predictions were created for each of the 24 subpoints. While the developed framework supports allocating an independent validation data set for the final model metric assessment, a single validation set is unlikely to provide useful insights due to the variability in the metrics for the small data sets in question. Allocating multiple validation data sets and running them through QPU was not feasible with the available time and resources, however; therefore, such a validation step was not included.

*Variability and errors*

For the classical models, Fig. 8. illustrates the distribution of balanced accuracy values for 200 different train/test splits for the configuration space coordinate with 300 samples and 10 features.

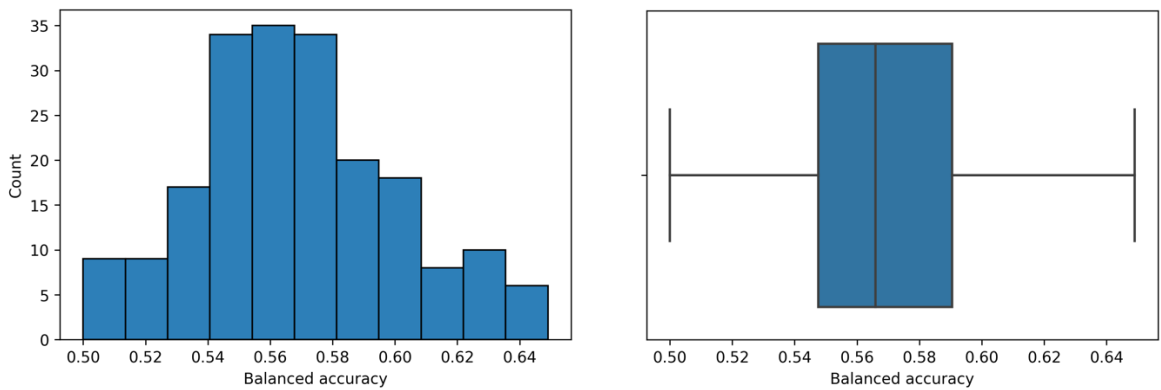

Fig. 8. Distribution of balanced accuracy values for the classical models for different train/test splits for the point with 300 samples and 10 features.

The scope of the SVM modeling was predicting the labels in the binary classification. We used scikit-learn and its predict method to predict labels directly and the predict_proba method for predicting approximate probabilities. Probabilities are estimated inside the predict_proba method using five-way cross-validation. As such, the process is subject to variability depending on the random seed that is provided to the svm.SVC call. One random value was used for calculations and comparisons to support reproducibility. In addition, a range of different random seed values was used in selected cases to obtain the variability of the predictions. It showed a 1.5 % standard deviation in the predictions obtained with the predict_proba method, both in the classical and quantum case, with less distinct values in the quantum domain.

*Runtimes*

The QPU runtimes were between 12 and 24 hours. During the QPU runs, both CPU and memory utilization was very low in the same Linux server as the main processing was executed on the QPU instance. The QPU processing was sequential.

The QPU processing was executed from a c5.18xlarge Ubuntu 18.04 cloud instance on Amazon Web Services (AWS), with 72 vCPUs and 144 GB Ram. We have not used Amazon Braket; instead, to support managing the whole process, we developed a Python package that has a code generation layer to simplify execution of different configuration spaces and management of the results. From within the package, the Qiskit API calls are made to simulators or QPUs. We used virtual environments that, over the course of a year, allowed us to effectively manage different versions of Qiskit and related software components.

*SHAP (SHapley Additive exPlanations)*

SHAP (SHapley Additive exPlanations) is a game-theoretic approach to explain the output of any machine learning model. It connects optimal allocation with local explanations using the classic Shapley values from game theory and their related extensions. SHAP measures the impact of variables by considering the interaction with other variables. Shapley values calculate the importance of a feature by comparing what a model predicts with and without the feature (variable)[32,33].

A starting list of the topmost 20 features was obtained from the original 16000-sample data sets by training machine learning models and then using SHAP to obtain the top 20 features. Those 20 features were then used in the aforementioned analysis with reduced size subsets in the order of SHAP relevance. Each of the points for 5, 10, 15, and 20 features on the configuration space was obtained by taking that many top features from the full 20-feature list, preserving the order.

*PTRI – Systematic identification of problems where quantum kernels may have empirical advantages*

Consider a set of classification problems where the number of features is between 1 and M and the number of samples between 1 and N. There are thus M x N classification problems. In order to help address the question which subset of these problems should be solved with a quantum kernel, we created a geophysics-inspired approach to identify regions of potential EQA in data sets. One way to select a suitable subset of problems involves studying the ruggedness of the manifold via PTRI, a metric we have adapted from[34] and defined as follows. We are considering the F1 score only as an instance of a metric in the formula. In the M X N configuration space defined before, each point is surrounded by eight other points except for the boundary points. For the boundary points, the performance result (F1 score in this case) of the adjacent points that are beyond the boundary is assumed to be 0. For the (i,j)-th interior point, the local $PTRI_{i,j}$ is calculated according to $PTRI_{i,j} = [(F1_{i,j} - F1_{i-1, j-1})^2 + (F1_{i,j} - F1_{i-1, j})^2 + (F1_{i,j} - F1_{i-1, j+1})^2 + (F1_{i,j} - F1_{i, j-1})^2 + (F1_{i,j} - F1_{i, j+1})^2 + (F1_{i,j} - F1_{i+1, j-1})^2 + (F1_{i,j} - F1_{i+1, j})^2 + (F1_{i,j} - F1_{i+1, j+1})^2]^{1/2}$. To determine the PTRI of the full configuration space, we average across the $PTRI_{i,j}$ values.

## Data availability

The data used in this work is commercially available via the Optum® de-identified Electronic Health Record data set.

## Code availability

The code was developed with Python 3 and Qiskit. Code sections may be obtained on request from corresponding author Z.K.

## Acknowledgements

The authors would like to thank Travis L. Scholten for helpful discussions and Winona Murphy, Matthew Stypulkoski, and Sarah Sheldon for supporting the dedicated access to ibmq_dublin.

## Competing Interests

The authors declare no competing interests.

## Author Contributions

Z.K. implemented the end-to-end computational pipeline, including the preprocessing of EHR data and the development of the quantum-classical machine learning models. F.F.F. contributed to the data preprocessing and machine learning model design. N.E.N. supported the optimization and execution of the experiments run on QPUs. O.S. led the theoretical quantum algorithm analysis and helped implement the quantum-classical models. All authors contributed to the design of the experiments, analyzed the results, and wrote the final manuscript.

## Correspondence

Correspondence and requests for materials should be addressed to Z.K.